\documentclass[a4paper]{article}
\usepackage{graphicx}
\usepackage{float}
\usepackage[utf8]{inputenc}
\usepackage{geometry}
\usepackage{hyperref}
\usepackage{authblk}

\title{Strong gravitational lensing around Kehagias-Sfetsos compact objects surrounded by plasma}

\author[1]{Sudipta Hensh}
\author[1]{Jan Schee}
\author[2,3,4,5,6]{Ahmadjon Abdujabbarov}
\author[1]{Zden\v{e}k Stuchl\'{i}k}
\affil[1]{Research Centre for Theoretical Physics and Astrophysics, Institute of Physics in Opava, Silesian University in Opava, Bezručovo náměstí 13, CZ-74601 Opava, Czech Republic}
\affil[2]{Ulugh Beg Astronomical Institute, Astronomy St. 33, Tashkent 100052, Uzbekistan}
\affil[3]{Institute of Nuclear Physics, Ulugbek 1, Tashkent 100214, Uzbekistan}
\affil[4]{National University of Uzbekistan, Tashkent 100174, Uzbekistan}
\affil[5]{Tashkent Institute of Irrigation and Agricultural Mechanization Engineers, Kori Niyoziy, 39, Tashkent 100000, Uzbekistan}
\affil[6]{Shanghai Astronomical Observatory, 80 Nandan Road, Shanghai 200030, P. R. China}

\date{}                     
\newcommand{\diff}{\mathrm{d}}

\begin{document}

\maketitle

\begin{abstract}
We present the analysis how Hořava gravity  and plasma
influence the strong lensing phenomena around Kehagias-Sfetsos (KS) black holes. 
Using the semi-analytical Bozza method of strong lensing limit, we determine the multiple images, namely their separation S, and magnification R. 
We apply our calculations to the case of supermassive black hole having mass $M=6.5\times 10^{9}\textrm{M}_\odot$ and being at distance $d_0=16.8\textrm{Mpc}$ from observer corresponding to those observed in M87. We show that the sensitivity of image magnification, image separation, and shadow angular size on KS parameter $\omega$ and plasma parameter $k$ are of order from $1\%$ to $10\%$ for $R$ and $~\,16\%$ for $S$.
\end{abstract}

\section{Introduction}
Recent observations of Event Horizon Telescope (EHT) \cite{2019ApJ...875L...1E,2019ApJ...875L...2E,2019ApJ...875L...3E,2019ApJ...875L...4E,2019ApJ...875L...5E,2019ApJ...875L...6E} related to the M87 galaxy often a new area of observational astrophysics and its relations to models of astrophysical phenomena occurring in strong gravity confined with strong electromagnetic fields. 
For details of interesting effects related to magnetized black holes see~\cite{Stuchlik:2020rls} however, the strongest information is related to the black hole shadow, discussed for the first time by Bardeen~\cite{1972ApJ...173L.137C}, and later by many authors, for M87 and SgrA its relevance was discussed, e.g. in~\cite{2009IJMPD..18..983S,Bambi:2019tjh,Wielgus:2020zvj}.
On the other hand, in the close vicinity of the black hole horizon, multiple images of radiating objects could be observed due to the effects of strong gravitational lensing, enabling thus an efficient way of testing alternative models of gravity, as in such circumstances the predictions of alternative gravity can differ significantly from those of general relativity.
The strong gravitational lensing is one of the key  phenomena, predicted by General Theory of Relativity (GR). It plays the crucial role in our attempt to distinguish among many concurrent theories arising since the advent of GR and determine the parameters of astrophysical compact objects that are responsible for strong gravitational lensing phenomena. 

The most direct attempt to calculate strong lensing effects is to numerically solve geodesics equations and determine associate magnitude and angular image separation. The pure numerical solutions encounter its limits when trying to determine the deflection of high order images. 

Luckily, there is a semi-analytical method developed by Bozza (\cite{Bozza01},\cite{Bozza02}) that splits the integral for deflection angle into analytical part and the numerical parts. The analytical part represents the integral that diverges as photons approach the photon circular orbit and is expressed as the logarithmic function of radial coordinate, while the numerical part is the regular integral, that can be calculated numerically with required accuracy.

Using Bozza's Strong Lensing Limit (SLL) we determine strong lensing effects in the vicinity of Kehagias-Sfetsos black-hole surrounded by plasma. The motivation is straightforward. 

The Kehagias-Sfetsos (KS) (\cite{Kehagias09}) static spherically symmetric metric is a solution of non-relativistic Hořava gravity field equations (\cite{Horava09a},\cite{Horava10}), the theory proposed as a ultraviolet (UV) completion of gravity which introduces non-equality of space and time in the UV regime, i.e. regime of Planck scales, however, at infrared regime standard GR is recovered. Hořava gravity touches the realm of the quantum gravity and strong lensing analysis in the vicinity of KS can give constraints of parameters defining black hole spacetime and corresponding link between GR and quantum gravity.
In fact, there is a series of works treating the astrophysical phenomena in the vicinity of KS black hole~\cite{Hensh:2019ipu} and works related to KS naked singularities~\cite{Stuchlik:2014jua,Stuchlik:2014iia,2014CQGra..31s5013S,Vieira:2013jga}.

The astrophysical black-holes are not isolated objects, but rather surrounded by matter, quite often in form of plasma. The radiation interacts not only with background spacetime but also with the plasma a photon no longer move along null geodesics given by background metric but rather along geodesics of effective geometry (introduced in sections to follow) \cite{Synge60}. 

Along with effect of Hořava gravity on strong lensing phenomena we analyse the sensitivity of it on plasma distributed in spherically symmetric fashion around KS black hole.




The paper is organized as follows. In 
Sect.~\ref{geometry} we introduce KS and effective KS metrices along with general formula for deflection angle. In Sect.~\ref{Bozza} we resume the basic Bozza's SLL fromulas we use to analyze effects of Hořava gravity and plasma on deflection, the image magnification and image separation quantities are introduced here too. The results are presented in Sect.~\ref{Results} and in Sect.~\ref{Conclusions} we discuss and conclude the obtained results.

Throughout the paper we will use a space-like signature $(-,+,+,+)$, a system of units in which $G = c = 1$ and, we shall restore them when we shall need to compare our results with observational data. Greek indices run from $0$ to $3$, Latin indices from $1$ to $3$.

\section{Deflection angle in KS with Plasma\label{geometry}}
The line element for Kehagias--Sfetsos spacetime is given by
\begin{eqnarray}
ds^2 = - f(r) dt^2 + \frac{dr^2}{f(r)} + r^2 (d \theta^2 + \sin^2 \theta d \phi^2 ) \ ,
\end{eqnarray}
where
\begin{eqnarray}
f(r) \equiv 1 + r^2\omega\left(1-\sqrt{1+4/(\omega r^3)}\right) \ .
\end{eqnarray}
We calculate deflection angle in the field of Kehagias-Sfetsos black hole surrounded by plasma with refractive index $n(r)$. 
We consider an isotropic distribution of plasma given by
\begin{eqnarray} \label{refractiveindex}
n^2(r) = 1- \frac{k}{r} \ ,
\end{eqnarray}
where $k$ is the plasma parameter determined by the properties of plasma.
Photons interact with plasma and therefore no longer move along null geodesics of the background spacetime but rather, as can be shown, along null geodesics of effective geometry that in our case read
\begin{equation}
	\diff s^2=-\frac{f(r)}{n^2}\diff t^2 + \frac{1}{f(r)}\diff r^2 + r^2\diff\theta^2+r^2\sin^2\theta\diff\phi^2.\label{effective_g}
\end{equation} 
Due to spherical symmetry of the spacetime it is sufficient to consider, without loss of generality, motion in equatorial plane, i.e. there is $\theta=\pi/2$ along the geodesics of interest. The metric (\ref{effective_g}) then reads
\begin{equation}
	\diff s^2 = -\frac{f(r)}{n^2(r)}\diff t^2 +\frac{1}{f(r)}\diff r^2 + r^2 \diff\phi^2.\label{effective_g1} \ .
\end{equation}
%
%
%
For above effective KS geometry given in~(\ref{effective_g1}), let us define 
\begin{equation} \label{ABCparameter}
	A(r) \equiv \frac{f(r)}{n^2(r)} \ , \quad B(r) \equiv \frac{1}{f(r)} \ , \quad \textrm{and} \quad C(r) \equiv r^2 \ .
\end{equation}
From the null vector normalization condition
\begin{equation}
	\tilde{g}_{\mu\nu}k^\mu k^\nu = 0 \ ,
\end{equation}
where tilde refers to effective geometry given in~(\ref{effective_g1}) and from existence of constants of motion $L\equiv k_\phi$ and $E\equiv -k_t$ one derives formula for the value of subtend azimuthal angle of photon geodesics in the form
\begin{equation}
	\Delta\phi(r_0) =2\int_{r_0}^\infty \frac{d \phi}{dr} dr= 2\int_{r_0}^\infty\frac{l\diff r}{r^2 \sqrt{n^2 (r)-\frac{f(r)l^2}{r^2}}}.\label{deltaphi}
\end{equation}
Here, $l\equiv L/E$ is the impact parameter and $r_0$ is the distance of closest approach i.e. corresponding turning point in radial coordinate and it is a solution of equation
\begin{equation}
	n^2 (r) r^2- f(r)l^2=0.
\end{equation}
The deflection angle $\alpha$ is then given by expression
\begin{equation}
	\alpha(r_0)=\Delta\phi(r_0) - \pi.
\end{equation}
Integral (\ref{deltaphi}) is valid only for $l\geq l_{\gamma}$ where $l_\gamma$ is impact parameter of photon circular orbit in effective geometry, i.e. it is determined from conditions
\begin{equation}
	k^u=0 \quad\textrm{and}\quad \frac{\diff k^u}{\diff\lambda}=0 \ ,
\end{equation}
leading to following couple of equations
\begin{eqnarray}
	n^2 (r_\gamma) r^2- f(r_\gamma)l_\gamma^2 &=& 0 \ ,\\
	2 n (r_\gamma)\,n'({r_\gamma}) r_\gamma^3
	+2 f(r_\gamma) l_\gamma^2 -2 f'({r_\gamma}) r_{\gamma} l_\gamma^2 &=& 0 \ ,
\end{eqnarray}
where the prime represents the derivative with respect to $r$ and $r_\gamma$ is  the radius of the photon circular orbit.
\section{Bozza's Strong Lensing Limit\label{Bozza}}
Bozza in his paper \cite{Bozza02} developed general  method to construct strong lensing limit, i.e. the regime of photon impact parameter is close to impact parameter of photon circular orbit of given spacetime.
Here we summarize the results of this procedure. 

Let us define a variable $z$ which is mutually connected with $r$ by transformation 
\begin{equation}
    z\equiv \frac{A-A_0}{1-A_0} \label{zformula} \ ,
\end{equation}
where the index `0' refers to quantities evaluated at turning point $r_0$.

We can rewrite the integral~(\ref{deltaphi}) as
\begin{eqnarray} \label{eq:13}
\Delta\phi(r_0) = \int_0^1 R(z,r_0)f(z,r_0)dz \ ,
\end{eqnarray}
where
\begin{eqnarray}
R(z,r_0) &=& \frac{2 \sqrt{BA}}{CA'}(1-A_0)\sqrt{C_0} \ ,\\
f(z,r_0) &=& \frac{1}{\sqrt{A_0-\left((1-A_0)z+A_0\right)C_0/C}} \ .
\end{eqnarray}
The function $f(z,r_0)$ diverges for $z \rightarrow 0$ but the function $R(z,r_0)$ is regular for all values of $z$. Using Taylor expansion, the function $f(z,r_0)$ can be approximated as
\begin{eqnarray}
f(z,r_0) \sim f_0(z,r_0)&=&\left(\zeta z+ \chi z^2\right)^{-1/2} \ ,
\end{eqnarray}
with
\begin{eqnarray}
    \zeta &=&\frac{1-A_0}{C_0A'_0}\left(C'_0 A_0- C_0 A'_0\right),\\
    \chi &=&\frac{(1-A_0)^2}{2C_0^2 A'^3_0}\left[2C_0 C'_0A'^2_0+(C0 C''_0-2C'^2_0)A_0 A'_0-C_0 C'_0 A_0 A''_0\right] \ .
\end{eqnarray}
The integral~(\ref{eq:13}) can be divided into regular~($I_R(r_0)$) and divergent parts~($I_D(r_0)$) as follows,
\begin{equation}
\Delta\phi(r_0)=I_D(r_0)+I_R(r_0) \ ,
\end{equation}
where
\begin{eqnarray}
I_D(r_0)&=&\int_0^1 R(0,r_\gamma) f_0(z,r_0) dz \ , \\
I_R(r_0)&=&\int_0^1 [R(z,r_0)f(z,r_0)-R(0,r_\gamma)f_0(z,r_0)] dz \ .
\end{eqnarray}
If we solve regular and divergent part of integral separately and assume the closest approach distance $r_0$ to be very close to $r_\gamma$ we can write the expression for deflection angle as 
\begin{equation}
    \alpha(\theta)=-\bar{a}\log\left[\frac{\theta\,D_{OL}}{l_{\gamma}}-1\right]+\bar{b}.
\end{equation}
where $D_{OL}$ is the distance between observer and lens, $\theta=l/D_{OL}$ is the image angular position in the lens system. The parameters that spacetime geometry determines are calculated form formulas
\begin{eqnarray}
\bar{a}&=&\frac{R(0,r_{\gamma})}{2\sqrt{\chi_\gamma}},\\
\bar{b}&=&-\pi+b_R+\bar{a}\log\left(\frac{2 \chi_\gamma}{A_\gamma}\right)
\end{eqnarray}
with following definitions
\begin{eqnarray}
    R(0,r_\gamma)&=&\frac{2 A_0}{C_0\,A'_0}(1-A_\gamma)\sqrt{C_\gamma},\\
    \chi_\gamma&=&\frac{C_\gamma (1-A_\gamma)^2\left[C''_\gamma A_\gamma-C_\gamma A''_\gamma\right]}{2A^2_\gamma C'^2_\gamma} \ ,
\end{eqnarray}
where the index `$\gamma$' refers to quantities evaluated at photon orbit, i.e. at $r_\gamma$.
The parameter $b_R$ is calculated by numerical integration of finite definite integral
\begin{equation}
    b_R=I_R(r_\gamma) \ .
\end{equation}
%

When evaluating $b_R$, i.e. calculating integral (28) we need expression for $r=r(z)$. Rearranging (13) we get
\begin{equation}
    A(r)=(1-A_0)z+A_0 \equiv F(z) .\label{Aeq}
\end{equation}
For given $z$ we can solve this equation numerically, but one can convince himself that this function is very steep at $z\sim 1$. Fortunately function $A(r)$ is rational and equation (\ref{Aeq}) can be solved as polynomial equation in the case of $n^2$ being given by (\ref{refractiveindex}). Equation (\ref{Aeq}) can then be rewritten to read
\begin{eqnarray}
2 \omega^2 (1- F(z))  r^7 + (2k \omega^2 F(z)- 4 \omega^2) r^6 + \omega (1+ F^2(z) -2 F(z))  r^5 + 2 k  \omega (1- F(z)) F(z) r^4 + k^2 \omega F^2 (z) r^3 = 0
\end{eqnarray}
%


\subsection{Image separation and magnification}
The strong lensing tool becomes very useful when deflection angle is associated with observable quantities giving us connection between observations and parameters of the model. The key equation is the lens equation derived for case of strong deflection limit also by Bozza in \cite{Bozza02} considering the observer, source and lens are almost aligned reads
\begin{equation}
    \beta = \theta - \frac{D_{LS}}{D_{OS}}\Delta\alpha_nm \ ,
\end{equation}
where $\beta$ is the angular position of the source i.e. angel between the direction of the source and the optical axis,
$D_LS$ and $D_OS$ are the lens-source and observer-source distance.
The quantity $\Delta\alpha_m$ is the offset of deflection angle given by
\begin{equation}
    \Delta \alpha_m\equiv \alpha(\theta)-2\pi m \ ,
\end{equation}
where $m$ indicates the number of loops made by the photon around the black hole.
Angular position of the $m$th order image is
\begin{equation}
    \theta_m=\theta_m^0+\frac{l_\gamma\,H_m\left(\beta - \theta^0 _m\right)\,D_{OS}}{\bar{a}\,D_{LS}\,D_{OL}} \ ,
\end{equation}
where
\begin{equation}
    H_m = \exp\left(\frac{\bar{a}-2n\pi}{\bar{b}}\right) \quad \textrm{and} \quad
    \theta_m^0 = \frac{l_\gamma}{D_{OL}} (1+ H_m) \ ,
\end{equation}
with $\theta_m^0$ is the angular position of the image when $\alpha(\theta_m^0)=2\pi\,m$\ .

The image magnification is associated with the ratio between image angular position $\theta$ and angular position of the source $\beta$, it reads
\begin{equation}
    \mu_m=\left(\frac{\beta}{\theta}\frac{\partial\beta}{\partial\theta}\right)^{-1}_{\theta_n^0}.
\end{equation}
Applying strong lensing lens equation, the resulting magnification formula is given by equation
\begin{equation}
    \mu_m=H_m\frac{l^2_\gamma \left(1+H_m\right)D_{OS}}{\bar{a}\beta\,D^2_{OL}\,D_{LS}}.
\end{equation}

\subsection{Observable quantities}
In order to make comparison between model and observations, potentially observable quantities must be defined. Here we consider the two natural quantities, already introduced by Bozza \cite{Bozza02}, the angular separation among the source images of different order and the magnification. First, let us discuss  images angular separation.  Due to the fact that the higher is the order of the image the more they are indistinguishable from the limiting case of images of infinite order, the natural choice of the angular separation parameter $S$ is 
\begin{equation}
    S\equiv \theta_1 - \theta_\infty,\label{Sparam}
\end{equation}
it is the angular separation between first order image and the limiting image order. Second, let us discuss the second parameter,$R$, expressing comparison between different images magnification parameter. The magnification of the images exponentially decreases with increasing order of the image. It is reasonable to define $R$ as quantity relating magnification of the first order image with the sum of magnification contributions of second and higher order images, i.e.
\begin{equation}
    R\equiv\frac{\mu_1}{\mu_{2+}} \ ,    \label{Rparam}
\end{equation}
where
\begin{equation}
    \mu_{2+}\equiv\sum\limits_{i=2}^{\infty}{\mu_i} \ .
\end{equation}
Using formulae introduced in previous subsection following from lens equation, the observable parameters read
\begin{equation}
    S=\theta_\infty \,\exp\left[\left(\bar{b}-2\pi\right)/\bar{a}\right] \ ,
\end{equation}
and
\begin{equation}
    R=\exp\left(2\pi/\bar{a}\right) \ .
\end{equation}



\section{Results\label{Results}}
\begin{figure}[H]
    \centering
    \begin{tabular}{cc}
    \includegraphics[scale=0.6]{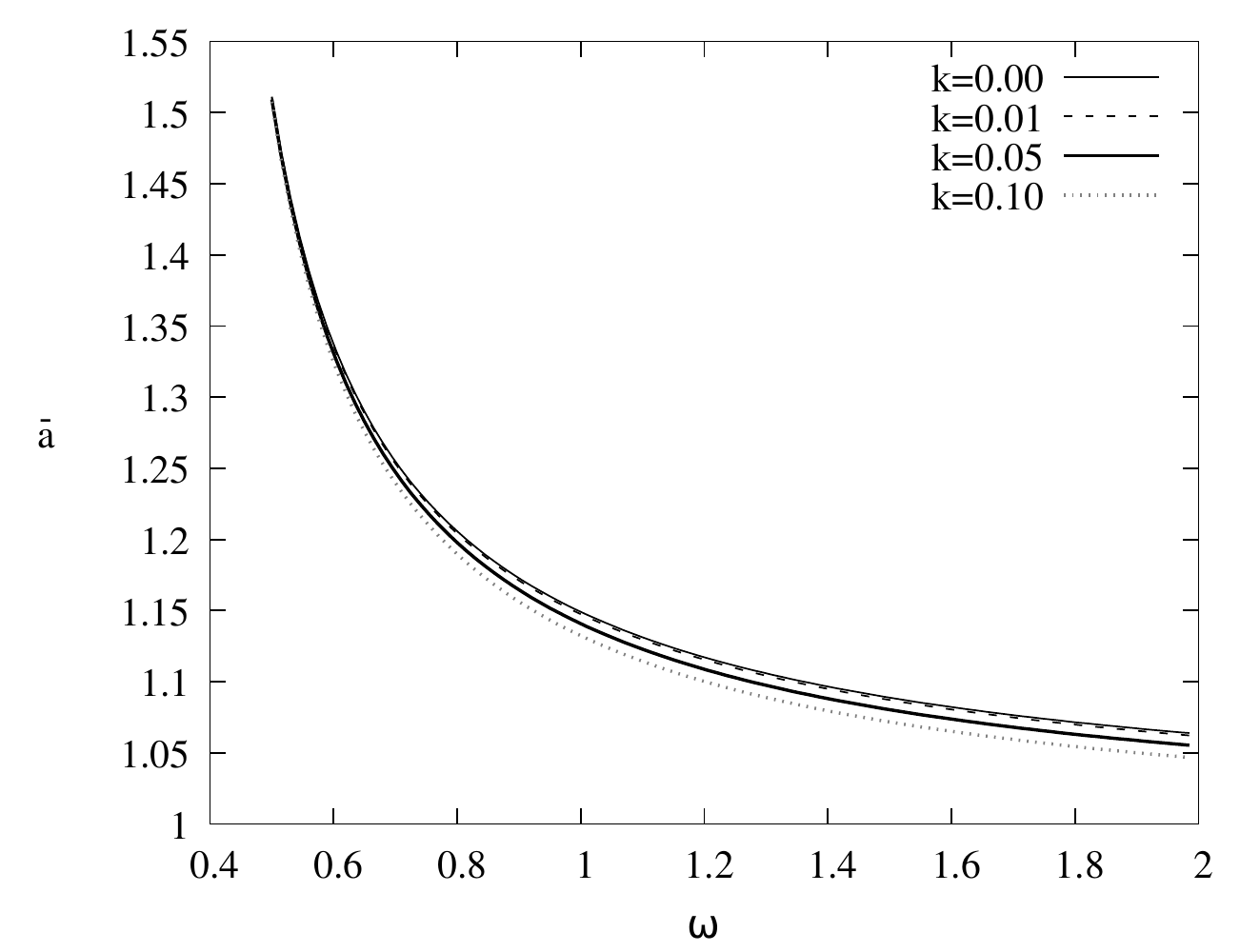} &\includegraphics[scale=0.6]{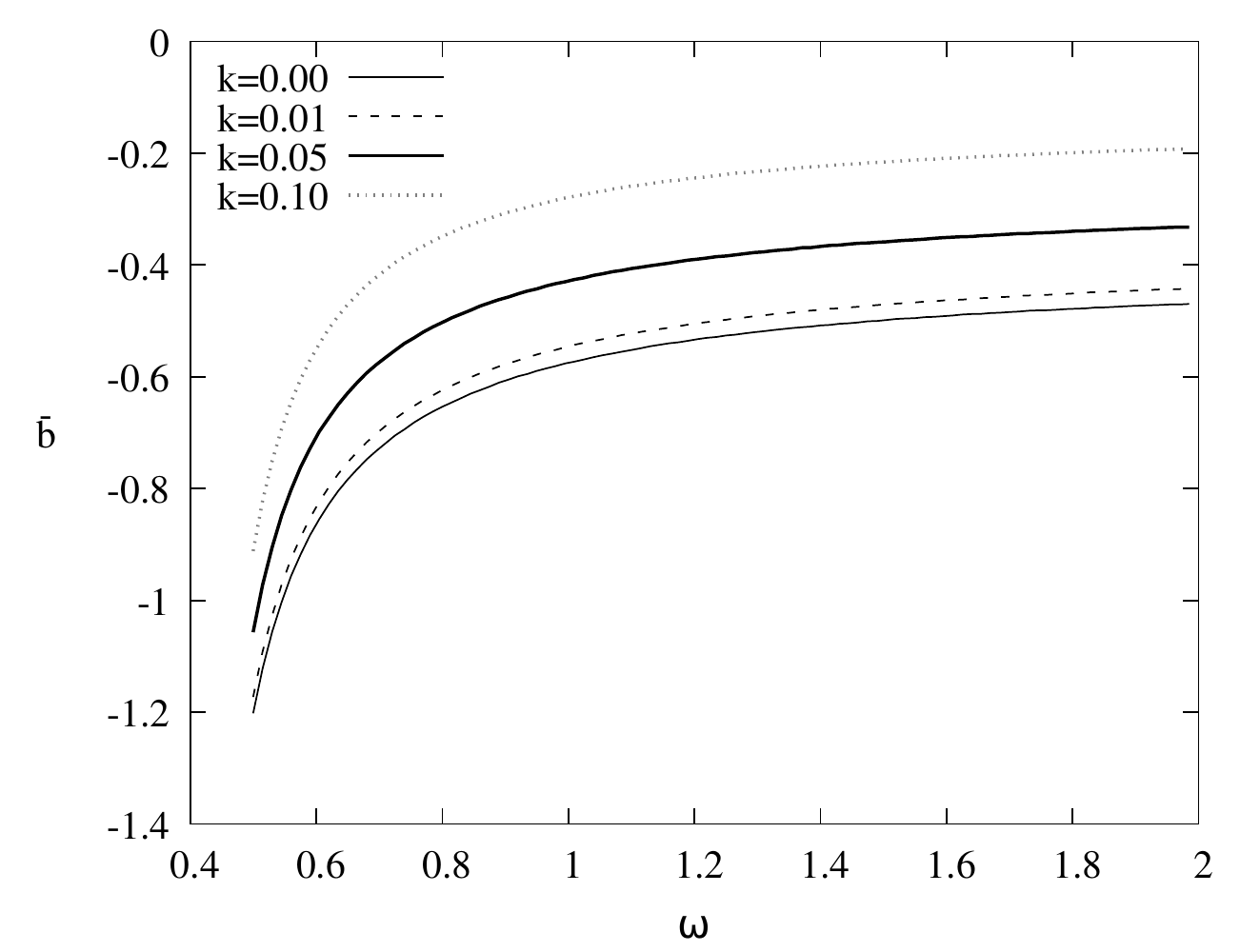}
    \end{tabular}
    \caption{Plots of $\bar{a}$ (left) and $\bar{b}$ (right) as functions of KS parameter $\omega\in[0.5,\,2.0]$ prepared for four representative values of plasma parameter $k=0.0$, $0.01$, $0.05$, and $0.1$ and plasma model (\ref{refractiveindex}). Note: the parameters $\bar{a}$ and $\bar{b}$ are in units of $rad$. }
    \label{fig_abar_bbar}
\end{figure}

\begin{figure}[H]
    \begin{center}
        \includegraphics[scale=0.35]{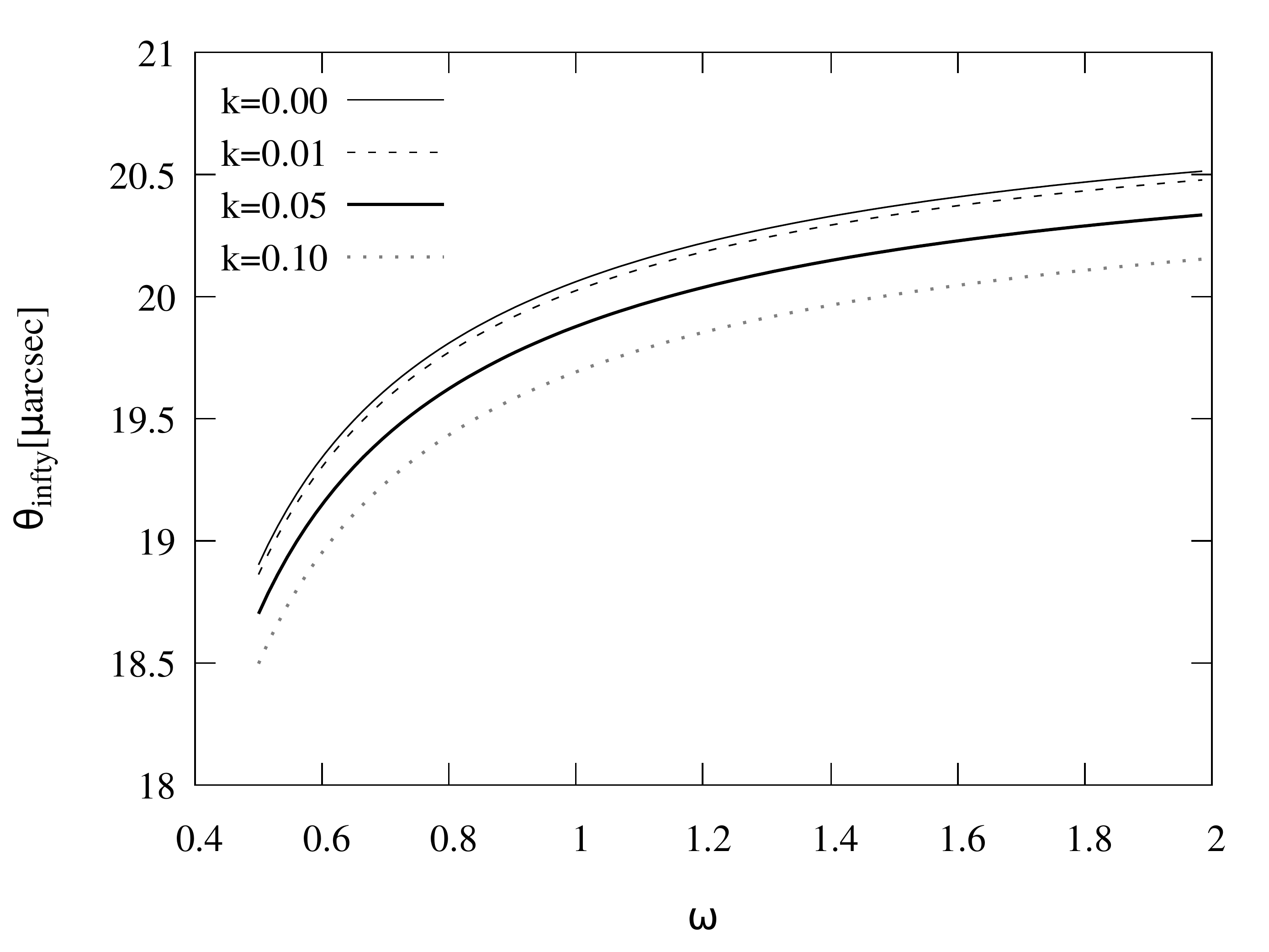}
        \caption{The black hole shadow angular size $\theta_\infty$ calculated for black-hole mass $M=6.5\times 10^9\,\mathrm{M}_\odot$ and distance $D_{OL}=16.0\mathrm{Mpc}$ (distance between lens and observer). \label{shadow}}
    \end{center}
\end{figure}

\begin{figure}[H]
    \centering
    \includegraphics[scale=0.35]{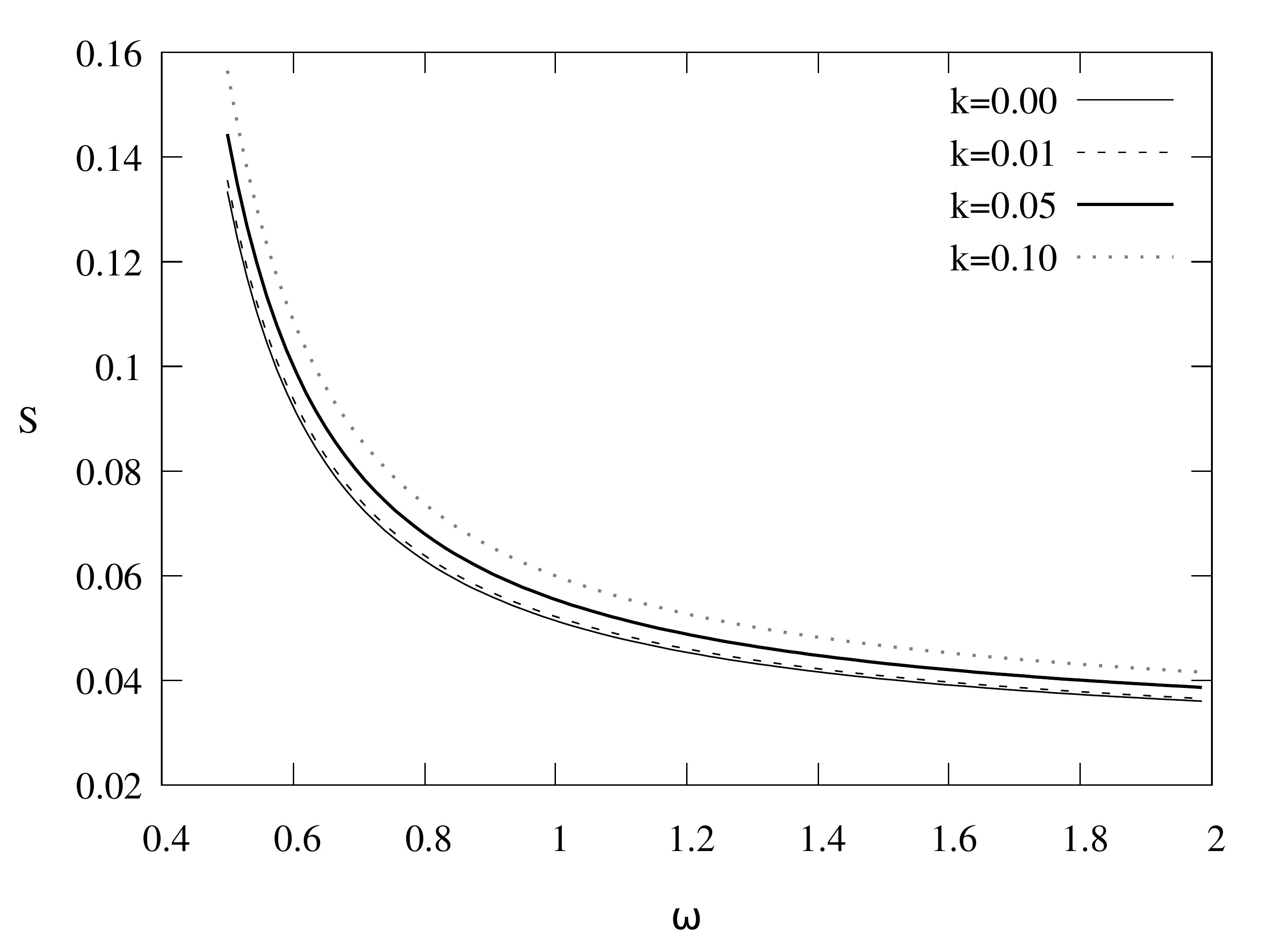}
    \caption{The images angular separation function $S$ plotted as function of KS parameter $\omega$ for four representative values of plasma parameter $k=0$,$0.01$, $0.05$, and $0.1$. The central lens mass is $M=6.5\times 10^9\,\mathrm{M}_\odot$  and distance $D_{OL}=16.0\mathrm{Mpc}$ (distance between lens and observer). }
    \label{figS}
\end{figure}

\begin{figure}[H]
    \centering
    \includegraphics[scale=0.35]{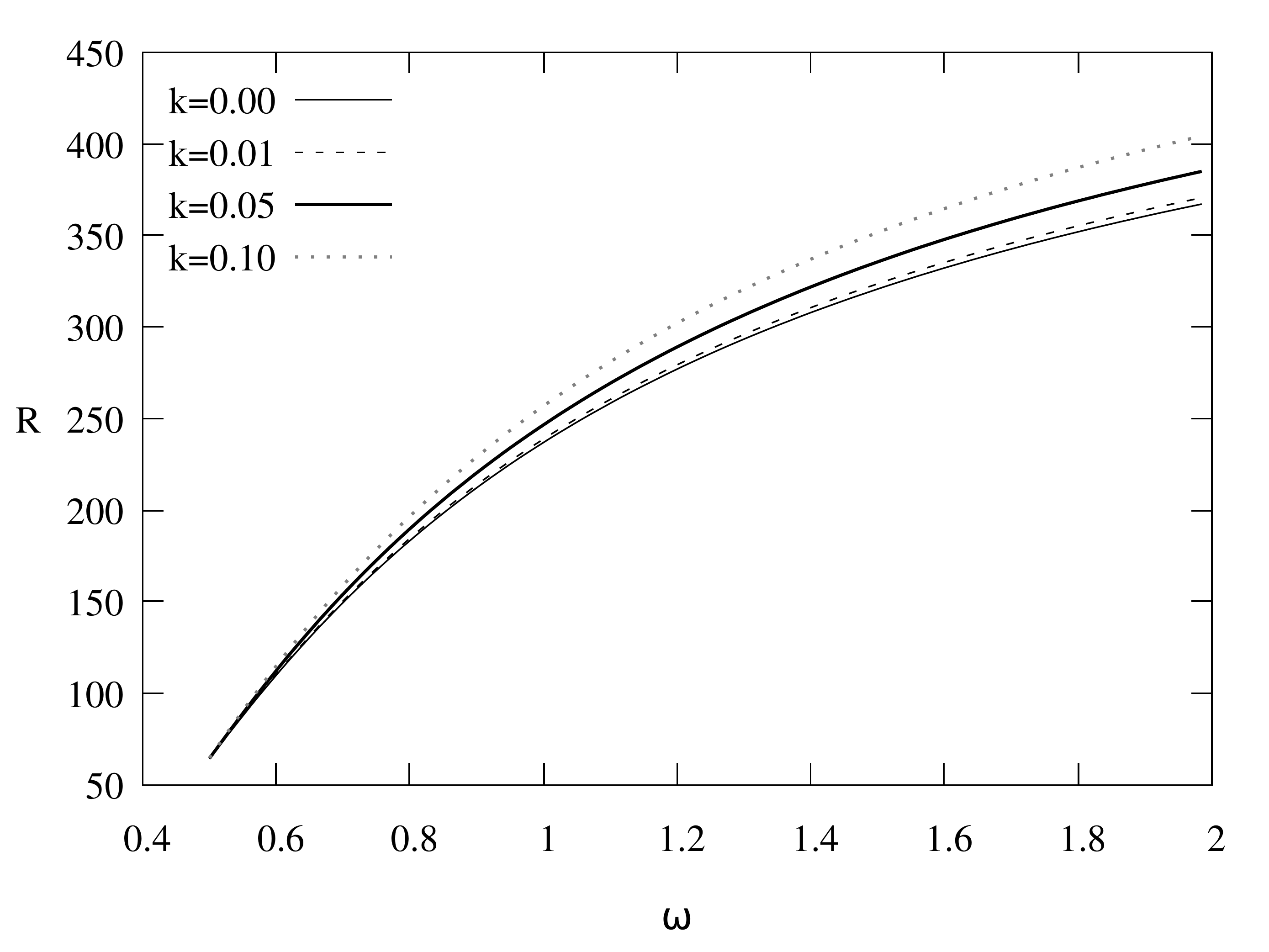}
    \caption{The images relative magnification function $R$ plotted as function of KS parameter $\omega$ for four representative values of plasma parameter $k=0$,$0.01$, $0.05$, and $0.1$. The central lens mass is $M=6.5\times 10^9\,\mathrm{M}_\odot$  and distance $D_{OL}=16.0\mathrm{Mpc}$ (distance between lens and observer). }
    \label{figR}
\end{figure}

The aim of this paper is to present and discuss the effect of both the KS spacetime and plasma on introduced observables $S$ and $R$. In our simulations we constrained ourselves to reasonable range of KS parameter  $\omega\in[0.5,\,2.0]$ and plasma model given in~(\ref{refractiveindex})
with parameter $k$ being picked up from set of representative values $\{0.0,\,0.01,\,0.05,\,0.1\}$. 
Both parameters depend on behavior of parameters $\bar{a}$ and $\bar{b}$, therefore we discuss the effect of plasma on those parameters (See Fig.~\ref{fig_abar_bbar}). Parameter $\bar{a}$ is always decreasing with increasing value of $\omega$ (recall, the higher is $\omega$ the closer is the KS spacetime to Schwarzchild spacetime).
The increasing plasma parameter $k$ increases the value of $\bar{a}$ making lensing effects stronger. On the other hand, the parameter $\bar{b}$ increases with increasing value of $\omega$ and the effect of plasma parameter $k$ is the same as in case of $\bar{a}$.  
Next important parameter determining the strong lensing effects if the angular size of the photon orbit, here represented by $\theta_{\infty}$ (see Fig.~\ref{shadow}). We clearly see that for fixed plasma parameter $k$ the KS shadow radius $\theta_{\infty}$ increases with increasing $\omega$ parameter on the interval $I_\omega\equiv[0.5,2]$. The increase of $\theta_{\infty}$ not linear. Increasing initial value of $\omega$ twice, the slope of  shadow radius increase is $2.28$ while for increase of initial value of $\omega$ four times, the slope of shadow angular radius increase is $1.07$. In all four considered cases of plasma parameter $k$, the shadow angular radius decreases and its relative change is   $2.1\%$ ($\omega=0.5$), $1.7\%$ ($\omega=1.0$), $1.75\%$ ($\omega=2.0$). 
In Figs.  \ref{figS} and \ref{figR} we present the effect of model on $S$ and $R$. We see that the angular separation $S$ is monotonically decreasing with increasing value of parameter $\omega$. We find that the relative change of $S$ between $k=0$ and $k=0.1$ is $\frac{\Delta S}{S_0}=17.3\%$ ($\omega=0.5$), $15.3\%$ ($\omega=1.0$), $15.6\%$ ($\omega=2.0$). In case of relative magnification $R$, its relative change between $k=0$ and $k=0.1$ is $\frac{\Delta R}{R_0}=0.95\%$ ($\omega=0.5$), $8.5\%$ ($\omega=1.0$), $10.1\%$ ($\omega=2.0$).

It is useful to present values of $R$, $S$, and $\theta_{\infty}$ for some representative values of $\omega$. We choose $\omega=0.5$, $1.0$, and $2$. The results are presented in table Tab.\ref{table1}.
\begin{table}[H]
    \centering
    \caption{The evaluation of function $R$, $S$ and $\theta_\infty$ at specific points $\omega$(spacetime) and $k$(plasma).The central lens mass is $M=6.5\times 10^9\,\mathrm{M}_\odot$  and distance $D_{OL}=16.0\mathrm{kpc}$ (distance between lens and observer). Here introduce $\tilde{R}\equiv 2.5\log R$.}\label{table1}
    \label{tab:my_label}
    \begin{tabular}{c|c|c|c|c|c}
    \hline
      $\omega\,M^{-2}$  & $k_i[1]$ & $R[1]$ & $\tilde{R}[\mathrm{mag}]$ & $S[\mu arcsec]$ & $\theta_\infty[\mu arcsec]$  \\
    \hline
      $0.5$     & $0.00$ & $63.949$ & $10.395$  & $0.133$     & $18.90$                     \\
                & $0.01$ & $64.023$ & $10.398$  & $0.136$     & $18.86$                     \\
                & $0.05$ & $64.291$ & $10.409$  & $0.144$     & $18.69$                    \\
                & $0.10$ & $64.555$ & $10.418$  & $0.156$     & $18.49$                 \\
  \hline
  $1.0$         & $0.00$ & $239.200$ & $13.690$  & $0.0515$    & $20.041$            \\
                & $0.01$ & $241.101$ & $13.713$ & $0.0517$    & $20.033$                 \\
                & $0.05$ & $248.971$ & $13.793$ & $0.0549$    & $19.886$             \\
                & $0.10$ & $259.446$ & $13.896$ & $0.0594$    & $19.700$                 \\
  \hline
  $2.0$         & $0.00$ & $367.09$  & $14.764$ & $0.0359$     & $20.512$          \\
                & $0.01$ & $370.55$  & $14.787$ & $0.0364$     & $20.477$                 \\
                & $0.05$ & $384.99$  & $14.883$ & $0.0385$     & $20.334$                \\
                & $0.10$ & $404.27$  & $15.005$ & $0.0415$     & $20.152$\\
\hline
    \end{tabular}
    
\end{table}
\newpage
\section{Conclusions\label{Conclusions}}
We applied Bozza's technique to Kehagias-Sfetsos black hole to analyze the effect of imprinted non-Lorentzian character of the Hořava gravity theory on strong lensing phenomena in the vicinity of the supermassive black-hole similar to one in M87 galaxy, i.e. the black hole mass  $6.5\times 10^9\,M_\odot$ and its distance is $16$kpc being surrounded with plasma. We modelled the environment around black hole by a plasma with refractive index $n$ (\ref{refractiveindex}). The model was controlled with  two parameters, $\omega$ representing Hořava gravity and $k$ representing the plasma and we studied the effect of them on astrophysically relevant quantities $S$ the angular separation parameter (\ref{Sparam}) and $R$ the magnification parameter (\ref{Rparam}). We have found out that for fixed value of parameter $k$ the value of $S$ parameter monotonically decreases with increasing parameter $\omega$ in range of $\omega>\omega_{mbh}$, while the value of $R$ parameter is monotonically increasing, on the same interval of $\omega$. Further, we have found out that the relative change of $S$ between $k=0$ and $k=0.1$ is $17.3\%$ ($\omega=0.5$), $15.3\%$ ($\omega=1.0$), $15.6\%$ ($\omega=2.0$). In case of relative magnification $R$, its relative change between $k=0$ and $k=0.1$ is $0.95\%$ ($\omega=0.5$), $8.5\%$ ($\omega=1.0$), $10.1\%$ ($\omega=2.0$).

We conclude, that the  effect of $\omega$ on angular separation $S$, in terms of $\Delta S/S_0$ decreases  with increasing $\omega$, but the change is not steep. On the other hand, there is increasing effect on relative magnification with increasing $\omega$ from $1\%$ for $\omega=0.5$ to $10\%$ for $\omega=2.0$. We have also calculate the effect of parameter $\omega$ on angular size of the black hole shadow. We see, from Table \ref{tab:my_label}, that increasing value of $\omega$ leads to the increase of $\theta_{\infty}$ with relative change between $k=0$ and $k=0.1$ being $\Delta\theta_\infty/\theta_\infty = 2.1\%$ ($\omega=0.5$), $1.7\%$ ($\omega=1.0$), and $1.7\%$ ($\omega=2.0$). The width of $\theta_\infty$ range is approximately $1.5\mu$arcsec which is too small to be a distinguishable effect today.

\section*{Acknowledgments}
We acknowledge the institutional support of Silesian University in Opava and Grant No. SGS/12/2019.   A. A. is supported by PIFI fund of Chinese Academy of Sciences.

\bibliographystyle{unsrt}  
\bibliography{ks_plasma_strong_lensing}

\end{document}